\documentstyle [12pt]{article}
\def\fnote#1#2{\begingroup\def\thefootnote{#1}\footnote{#2}\addtocounter
{footnote}{-1}\endgroup}

\def\inbar{\vrule height1.5ex width.4pt depth0pt}
\def\IB{\relax{\rm I\kern-.18em B}}
\def\IC{\relax\,\hbox{$\inbar\kern-.3em{\rm C}$}}
\def\ID{\relax{\rm I\kern-.18em D}}
\def\IE{\relax{\rm I\kern-.18em E}}
\def\IF{\relax{\rm I\kern-.18em F}}
\def\IG{\relax\,\hbox{$\inbar\kern-.3em{\rm G}$}}
\def\IH{\relax{\rm I\kern-.18em H}}
\def\II{\relax{\rm I\kern-.18em I}}
\def\IK{\relax{\rm I\kern-.18em K}}
\def\IL{\relax{\rm I\kern-.18em L}}
\def\IM{\relax{\rm I\kern-.18em M}}
\def\IN{\relax{\rm I\kern-.18em N}}
\def\IO{\relax\,\hbox{$\inbar\kern-.3em{\rm O}$}}
\def\IP{\relax{\rm I\kern-.18em P}}
\def\IQ{\relax\,\hbox{$\inbar\kern-.3em{\rm Q}$}}
\def\IR{\relax{\rm I\kern-.18em R}}
\def\ZZ{\relax{\sf Z\kern-.4em Z}}
\def\fnote#1#2{\begingroup\def\thefootnote{#1}\footnote{#2}\addtocounter
{footnote}{-1}\endgroup}
\def\beq{\begin{equation}}
\def\eeq{\end{equation}}
\def\bea{\begin{eqnarray}}
\def\eea{\end{eqnarray}}
\def\lleq#1{\label{#1}\eeq}
\def\llea#1{\label{#1}\eea}
\let\nn=\nonumber
\def\tabroom{\hbox to0pt{\phantom{\huge A}\hss}}
\def\notin{\ \hbox{{$\in$}\kern-.51em\hbox{/}}}
\def\a{\alpha}        
   \def\k{\kappa}  \def\l{\lambda}
     \def\si{\sigma}
   \def\th{\theta}
   
   \def\cI{{\cal I}}
\def\cJ{{\cal J}}   
   
\def\cR{{\cal R}}

 \def\tM{\tilde M} \def\tV{\tilde V}

\def\lra{\longrightarrow}
\def\lolra{\longleftrightarrow}
  
\def\del{\partial}

\def\hbar{\bar h}

\def\ags{h^{(1,1)}}
\def\gens{h^{(2,1)}}

\begin{document}
\baselineskip=15pt
\parskip .1truein

\hfill {hep--th/9511058}
\vskip -.1truein
\hfill {NSF--ITP--95--141}
\vskip -.1truein
\hfill {BONN--TH--95--20}
\vskip .3truein
\noindent

\centerline{\large CONIFOLD TRANSITIONS AND MIRROR SYMMETRIES}

\vskip .4truein
\centerline{\sc M.Lynker$^1$
                   \fnote{\diamond}{Email: mlynker@siggy.iusb.indiana.edu}
                and
                R.Schimmrigk$^{2,3}$
                   \fnote{\dagger}{Email: netah@avzw02.physik.uni-bonn.de}
            }

\vskip .3truein
\centerline{\it $^1$ Department of Physics, Indiana University South Bend}
\vskip .01truein
\centerline{\it 1700 Mishawaka Ave., South Bend, Indiana, 46634}

\vskip .1truein
\centerline{\it $^2$ Physikalisches Institut, Universit\"at Bonn}
\vskip .01truein
\centerline{\it Nussallee 12, 53115 Bonn}

\vskip .1truein
\centerline{\it $^3$ Institute for Theoretical Physics,
         University of California}
\vskip .01truein
\centerline{\it Santa Barbara, California 93106}

\vskip 0.5truein
\centerline{\bf ABSTRACT}

\noindent
Recent work initiated by Strominger has lead to a consistent physical
interpretation of certain types of transitions between different string
vacua. These transitions, discovered several years ago, involve singular
conifold configurations which connect distinct Calabi-Yau manifolds. In
this paper we discuss a number of aspects of conifold transitions pertinent
to both worldsheet and spacetime mirror symmetry. It is shown that the
mirror transform based on fractional transformations allows an extension of
the mirror map to conifold boundary points of the moduli space of weighted
Calabi-Yau manifolds. The conifold points encountered in the mirror context
are not amenable to an analysis via the original splitting constructions. We
describe the first examples of such nonsplitting conifold transitions, which
turn out to connect the known web of Calabi-Yau spaces to new regions of the
collective moduli space. We then generalize the splitting conifold transition
to weighted manifolds and describe a class of connections between the webs of
ordinary and weighted projective Calabi-Yau spaces. Combining these two
constructions we find evidence for a dual analog of conifold transitions
in heterotic N$=$2 compactifications on K3$\times $T$^2$ and in particular
describe the first conifold transition of a Calabi-Yau manifold whose
heterotic dual has been identified by Kachru and Vafa. We furthermore
present a special type of conifold transition which, when applied to certain
classes of Calabi-Yau K3 fibrations, preserves the fiber structure.

\renewcommand\thepage{}
\newpage

\baselineskip=17.5pt
\parskip=.1truein
\parindent=20pt
\pagenumbering{arabic}

\noindent
{\bf 1. Introduction}

\noindent
The fact that singular varieties of conifold type describe transition points
between Calabi-Yau manifolds with different Hodge numbers has been realized
several years ago in the framework of complete intersection Calabi-Yau
manifolds embedded in products of projective spaces \cite{cdls88}.
Only recently however has a physical interpretation of such transitions
been found through work initiated by Strominger \cite{as95}.
The picture which emerges is that the singular
configurations of \cite{cdls88} correspond, in the low--energy
effective action, to divergences
that arise from integrating out massive modes which become massless at the
singularity.  The question whether the string can propagate consistently on
conifold configurations is important not only because of the possibility of
phase transitions between Calabi-Yau manifolds in the early universe, but
also because it is of relevance for problems in mirror symmetry \cite{ms},
and heterotic K3$\times $T$^2$-type II Calabi-Yau duality
\cite{kv95,fhsv95}, so--called worldsheet and spacetime mirror symmetry,
or first and second quantized mirror symmetry \cite{fhsv95}.
Indeed, it was remarked in \cite{gms95} that the link provided by the
splitting process between the individual moduli spaces of many
Calabi-Yau manifolds should make it possible to
extend mirror symmetry to the global moduli space of all Calabi-Yau spaces.
It was furthermore emphasized in \cite{kv95,fhsv95} that the
conjectured heterotic-type II duality inevitably leads
to the problem of singular
configurations because the moduli space of K3$\times $T$^2$ contains
points of enhanced symmetries where the massless spectrum changes.

Conifold transitions between Calabi-Yau spaces were introduced
in \cite{cdls88} in the context of complete intersection manifolds
embedded in products of ordinary projective space, and subsequent
discussions \cite{gh88, cgh90} have focused exclusively on such manifolds.
In the intervening years however this class has been found to be wanting.
Both, worldsheet and spacetime mirror
symmetry, necessitate a more general framework which, in first
approximation, is provided by weighted projective varieties\fnote{1}{We
will not discuss manifolds embedded in toric varieties.}.
One of the virtues of the class of weighted complete intersection
Calabi-Yau manifolds lies in the fact that for a large subclass of these
spaces we have at our disposal a physical construction of mirror
pairs\cite{ls90}, based on fractional transformations.
After reviewing and extending the discussion
of the mirror transform of \cite{ls90} in Section 2
we show in Section 3 that via this construction it is possible to
explicitly mirror map appropriate submanifolds of the moduli space of
Calabi-Yau spaces. These submanifolds of moduli space generically
contain conifold points and therefore the fractional mirror transform
allows us to trace Calabi-Yau configurations
to the boundary of moduli space, thereby providing a concrete
realization of the scenario envisioned in \cite{gms95}.
The singular varieties encountered in these regions of
moduli space are, however, not of the type originally discussed in
\cite{cdls88}, and therefore might be expected to have novel features.
In Section 4 we describe the first examples of such conifold transitions
relevant for mirror symmetry and find that indeed the resolved manifolds
open paths into new regions of the collective moduli space of all
Calabi-Yau manifolds.

Section 5 leads us from world sheet mirror symmetry to
heterotic-type II duality. We describe a construction, based again on
fractional transformations, which provides connections between various
subwebs of Calabi-Yau moduli spaces. The classes of manifolds
involved turn out to consist of spaces which are all K3 fibrations,
and therefore are of relevance for a second unification problem -
spacetime mirror symmetry. Part of our later discussion will therefore
focus on these classes of spaces.
Even though the particular types of conifold transitions described
in \cite{cdls88}, so--called splitting and contraction, are not
the most general kind of transition involving nodes, as we emphasize
in Section 4, they do have an important advantage. Whereas
in general it can be quite difficult to decide whether the resolved
manifold does in fact define a Calabi-Yau manifold, rather than some
more general space, the splitting construction not only automatically
guarantees that the resolved manifold is of Calabi-Yau type, it
also provides an explicit algebraic representation of the two
Calabi-Yau manifolds connected through some common conifold
configuration. This simplifies the analysis of the whole process
considerably. Because, as just mentioned, both types of mirror symmetry
 demand the class of weighted manifolds as a sort of
`minimal' framework, it is clearly of importance to generalize the
discussion of \cite{cdls88} to this more general context.
Perhaps the most intriguing problem is the possibility of finding a
heterotic analog of the Calabi-Yau conifold transition.

In the remaining part of the
 paper we initiate the analysis of conifold transitions
of splitting type in the context of weighted complete
intersection Calabi-Yau manifolds. The generalization of the
constructions of \cite{cdls88} to the weighted framework introduces
some new twists which we discuss in the two final Sections.
The first problem which has to be circumvented
concerns the question of transversality of the split configuration.
As in the case of weighted hypersurfaces it is not always possible
to find a quasismooth manifold for a given combination of
weights. We deal with this question in Section 6 and describe a
class of weighted splits which do connect quasismooth varieties.
In our discussion of such weighted conifold transitions
we will find  support for the speculation in \cite{kv95} that a
heterotic analog of conifold transitions indeed exists.
In particular we describe the first conifold transition of a
manifold whose heterotic dual has previously been suggested by Kachru
and Vafa \cite{kv95}\fnote{2}{A different type of transition
      has been discussed in \cite{pa95}.}.
The generalization of the splitting construction of \cite{cdls88} to the
weighted framework also provides support for the notion of a
universal moduli space of Calabi-Yau manifolds, generalizing to the
weighted category observations made in \cite{cdls88,gh88} in the
context of ordinary complete intersection Calabi-Yau spaces.

Finally, we describe the behavior of the K3 fibration of Calabi-Yau
manifolds, introduced in Section 5, under conifold transitions.
The fact that K3 fibrations are central to the problem of
heterotic-type II duality has been recognized in \cite{klm95}
and further discussed in \cite{al95}. We introduce a particular
type of splitting type conifold transition which preserves the
K3 fibration structure, thereby showing that the property
of K3 fibrations for Calabi-Yau threefolds extends to the class of
general complete intersection manifolds of arbitrary codimension,
and therefore is much more general than hitherto expected.
We end by describing in Section 7 the new phenomenon of
`colliding singularities'
which occurs in conifold transitions between weighted manifolds.

\vskip .2truein
\noindent
{\bf 2. The Fractional Transformation Mirror Transform}

\noindent
Our main tool in tracing the mirror map along certain directions to the
boundary of moduli space is the mirror transform based on fractional
transformations.  This construction was introduced in \cite{ls90} in order
to establish explicitly
the existence of mirror symmetry discovered in the first reference of
\cite{ms} in the framework of weighted Calabi-Yau manifolds.
In the following we briefly review the discussion of \cite{ls90}
and make it more precise\fnote{3}{Due to some mishap
      this article has appeared twice. The paper
     published in Phys.Lett. {\bf B268}(1991)47 is an identical copy of
      \cite{ls90}.}.

The essential ingredient of the fractional transformation mirror construction
is the basic isomorphism\fnote{4}{In \cite{ls90} the modding
     on the rhs of this relation was ignored because in all the
    applications discussed in that paper this
      additional orbifolding in the image theory turned out to be trivial
     simply because the action became part of the projective equivalence.
    In general, however, the action on the rhs can not be neglected.}
\bea
& &\IC_{\left(\frac{b}{g_{ab}},\frac{a}{g_{ab}}\right)}
                                            \left[\frac{ab}{g_{ab}}\right]
    \ni \left \{z_1^a+z_2^b=0\right \}
                 ~{\Big /}~ \ZZ_b: \left[\matrix{(b-1)&1}\right] ~~
                                              \nn \\ [3ex]
&\sim & \IC_{\left(\frac{b^2}{h_{ab}},\frac{a(b-1)-b}{h_{ab}}\right)}
                                         \left[\frac{ab(b-1)}{h_{ab}}\right]
       \ni \left \{y_1^{a(b-1)/b}+y_1y_2^b=0\right\}
                 ~{\Big /}~ \ZZ_{b-1}: \left[\matrix{1&(b-2)}\right]
\llea{basic-iso}

\noindent
induced by the fractional transformations
\bea
z_1 = y_1^{1-\frac{1}{b}}, & & y_1=z_1^{\frac{b}{b-1}} \nn \\
z_2=y_1^{\frac{1}{b}}y_2,  & & y_2=\frac{z_2}{z_1^{\frac{1}{b-1}}}
\eea
in the path integral of the theory.
Here $g_{ab}$ is the greatest common divisor of $a$ and $b$ and
$h_{ab}$ is the greatest common divisor of $b^2$ and $(ab-a-b)$.
The action of a cyclic group $\ZZ_b$ of order $b$ denoted by
$[m~~n]$ indicates that the symmetry acts like
$(z_1,z_2) \mapsto (\a^m z_1, \a^n z_2)$ where $\a$ is the $b^{th}$ root
of unity.

The ideal of the first cover theory
\beq
\cJ_z = \left[\del_1 p, \del_2 p \right]= \left[z_1^{a-1},z_2^{b-1}\right].
\eeq
generates the $\mu=(a-1)(b-1)$--dimensional ring
\beq
\cR_z = \{ z_1^pz_2^q ~|~0\leq p\leq a-2,~~0\leq q\leq b-2\}
\eeq
whereas the ideal of the second cover theory
\beq
\cI_y = \left[\del_1 p, \del_2 p \right] =
      \left[\frac{a(b-1)}{b} y_1^{\frac{a(b-1)}{b}-1} +y_2^b,
y_1y_2^{b-1}\right]
\eeq
generates the $\frac{a}{b}(b-1)^2+1$--dimensional polynomial ring
\beq
\cR_y = \{y_2^{b-1}\} \cup
 \{y_1^py_2^q~|~0\leq p\leq \frac{a}{b}(b-1)-1, 0\leq q \leq (b-2) \}
\eeq

We are interested in the states of the orbifold theories. First consider
the invariant sectors:
\beq
\cR_z^{inv} = \{ z_1^pz_2^q ~|~0\leq p\leq a-2,~~0\leq q\leq b-2,~~
                               p(b-1)+q =0 ~~{\rm mod}~~b\}
\eeq
i.e. $p-q=0$ mod $b$ and therefore $p=q+nb$ for some integer $n\in \IN$.
Thus dim~$\cR_z^{inv} = a(b-1)/b$.  Similarly
\beq
\cR_y^{inv}= \{y_2^{b-1}\} \cup
 \{y_1^py_2^q~|~0\leq p\leq \frac{a}{b}(b-1)-1, 0\leq q \leq (b-2) ,
                  p+q(b-2) =0 ~~{\rm mod}~~(b-1)\}
\eeq
and hence $p=q+n(b-1)$ and the dimension is
dim$~\cR_y^{inv} = a(b-1)/b+1$. Hence there is only one twisted state in the
$z$--orbifold which is mapped by fractional transformations into a monomial
of the $y$--theory.

It follows from the analysis in \cite{ls90} that twisted states are of the
form $\left(z_2^p/z_1^q\right)$ with $p,q\in \ZZ$. The first constraint comes
from invariance under the $\ZZ_b$ action,
which leads to the relation $p=nb+q(b-1)$ for some integer $n\in \ZZ$.
Thus these rational forms take the form
$z_2^{nb+q(b-1)}/z_1^q \leftrightarrow y_1^ny_2^{nb+q(b-1)}$ , for
$n,q \in \ZZ$ with the unitarity constraint
$q\left(1 -1/a - 1/b\right) + n \geq 0$. For $n\geq 1$, $q\geq 0$
 clearly all image states are in the ideal of the $y$--theory.
Hence the states above with $q> 0$, any $n$, and $q< 0, nb < -q(b-1)$
are possible twisted states, which, for $n=0,q>0$ lead to monomials:
$z_2^{q(b-1)}/z_1^q \longleftrightarrow y_2^{q(b-1)}$.
For $q>2$ the $y$--monomials belong to the ideal as well, leaving us with
two twisted states $z_2^{b-1}/z_1 \leftrightarrow y_2^{b-1}$
and $z_2^{2(b-1)}/z_1^2 \leftrightarrow y_2^{2(b-1)}$.
Both of these states are in the invariant sector of the image theory and
thus survive the $\ZZ_{b-1}$--modding. The final reduction comes from
realizing that the state $z_2^{2(b-1)}/z_1^2 \leftrightarrow y_2^{2(b-1)}$
is in fact equivalent to an invariant state: via the $y$--ideal the above state
can be written as $y_2^{b}y_2^{b-2} = y_1^{\frac{a}{b}(b-1) - 1}y_2^{b-2}$
which the fractional transformation map into $z_1^{a-2}z_2^{b-2}$
which is the top state in the $\cR_z$ ring invariant with respect to the
$\ZZ_b$--action.
Thus the invariant sector of the $y$--orbifold theory is mapped into
the invariant ring of the $z$--theory plus one twisted state.
A simple application of this discussion to the isomorphism
$\IP_{(1,7,2,2,2)}[14]\sim \IP_{(1,3,1,1,1)}[7]$ \cite{ls90} allows an
explicit relation between the purely polynomial chiral ring on the rhs and the
chiral ring on the lhs which is supplemented by blow--up modes originating
from the singular set described by a genus 15 curve. The blow--up modes thus
acquire a representation as rational expressions in the coordinates.

The basic isomorphism itself provides the mirror of weighted spaces only in
very few cases, such as $\IP_{(3,8,33,66,88,132)}[264]^{(57,81)}/\ZZ_2
   \sim \IP_{(3,8,66,88,99)}[264]^{(81,57)}$.
Much more powerful, however, is a simple iteration of the basic
isomorphism as described in \cite{ls90}.

It is important to realize that even though the basic isomorphism maps a
Fermat type orbifold into a tadpole orbifold the fractional transformation
mirror transform is not restricted to Fermat type polynomials.
Consider e.g. the manifold embedded in
\beq
\IP_{(3,6,6,4,5)}[24]^{(10,34)}_{-48}
\ni \{ p= z_1^8 + z_2^4 + z_3^4 + z_4^6 + z_4 z_5^4=0\}.
\eeq
Orbifolding this space with respect to a $\ZZ_4$ symmetry  with the action
$\ZZ_4~:~ [~0~0~1~0~3~]$ and using the appropriate
fractional transformations as discussed above leads to the mirror manifold
\beq
\IP_{(9,18,12,13,20)}[72]^{(34,10)}_{48}
\ni \{p= z_1^8 + z_2^4 + z_3^6 + z_3z_5^3 + z_5z_4^4=0\}.
\eeq

\vskip .2truein
\noindent
{\bf 3. Mirror Mapping Moduli Spaces}

\noindent
We can now apply fractional transformations to map moduli spaces. A
simple example which illustrates this phenomenon is furnished by the quintic
which, at the exactly solvable point, takes the form
\beq
(3^5)_{A_4^5}~~ \sim ~~\IC^*_{(1,1,1,1,1)}[5]~~ \sim ~~
\IP_4[5]\ni \left\{\sum_{i=1}^5 z_i^5=0\right\}.
\eeq
Using the iteration of the basic isomorphism as described in Section 2 one
finds that orbifolding the Landau--Ginzburg theory with respect to
the cyclic group
\beq
\ZZ_5^3:~~\left[\matrix{4 &1 &0 &0 &0\cr
                        0 &4 &1 &0 &0\cr
                        0 &0 &4 &1 &0\cr}\right]
\lleq{quintact}
leads to the mirror spectrum. Using the fractional transformations
\beq
z_1=y_1^{4/5},~z_2=y_1^{1/5}y_2^{4/5},~z_3=y_2^{1/5}y_3^{4/5},z_4=y_3^{1/5}y_4,~
z_5 = y_5
\eeq
that follow from the iteration of the basic isomorphism shows that this
orbifold is isomorphic to the complete intersection
\beq
\IP_{(80,60,65,51,64)}[320]_{200}^{101} \ni
              \{z_1^4+z_1z_2^4+z_2z_3^4+z_3z_4^5+z_5^5=0\},
\lleq{quintmirr1}
where again no orbifolding is necessary on the mirror side of the isomorphism.
The manifold (\ref{quintmirr1}) is indeed one of two spaces in the list of
weighted hypersurfaces which has the appropriate mirror spectrum of the quintic
(see the first ref. of \cite{ms} and \cite{all-lgs}).

It is important to emphasize that the fractional mirror transform
can be applied to the whole relevant part of the moduli space: since the
complex sector of the mirror of the quintic is 1--dimensional, we need to
establish a map that relates a 1--dimensional subspace of the 101--dimensional
space of complex
deformations to the 1--dimensional subspace of the mirror.
This is achieved by considering
\beq
\IP_4[5]\ni \left\{\sum_{i=1}^5 z_i^5-5\psi \prod_{i=1}^5 z_i=0\right\}.
\lleq{quint-def}
The crucial point here is that for each value of $\psi$ the configuration
features the mirror discrete group (\ref{quintact}) as a symmetry group and
we can mod out this action. Furthermore for each value of $\psi$ we can
apply our fractional transformation and map this configuration into
\beq
\IP_{(80,60,65,51,64)}[320]_{200}^{101} \ni
                                 \{z_1^4+z_1z_2^4+z_2z_3^4+z_3z_4^5+z_5^5
              -5\psi \prod_{i=1}^5 z_i=0\}.
\lleq{quintmirr1mod}
The same holds for the second representation of the mirror which appears
in the lists of \cite{all-lgs}
\fnote{5}{ The second configuration with the mirror spectrum of the quintic,
$\IP_{(64,48,52,51,41)}[256]_{200}^{101}$ is isomorphic to the one just
discussed because the additional $\ZZ_5$ modding via which it is obtained
from the quintic is part of the projective equivalence.}.

The above example of a map between moduli spaces is the simplest
example of a vast class of manifolds, the moduli spaces of which feature the
very same structure as the quintic in a
one--dimensional subspace. In the class of
weighted Calabi-Yau hypersurfaces of degree $d$ in weighted projective space
$\IP_{(k_1,k_2,k_3,k_4,k_5)}[d]$ with $d= \sum_{i=1}^5 k_i$
there always exists a 1--dimensional family of manifolds
\beq
\IP_{(k_1,k_2,k_3,k_4,k_5)}[d] \ni
\left\{ p_0(z_i) - d \psi \prod_i z_i =0 \right\}.
\lleq{1-dim-fam}
Since any action preserving the holomorphic threeform also leaves invariant
this canonical deformation and furthermore fractional transformations leave
invariant this monomial it follows without any further check that the
fractional mirror transform  always applies to this 1--dimensional subspace.
The important aspect of this
class is that by moving along this canonical direction in moduli
space one eventually runs into singularities, which generically are
conifolds. Thus our mirror construction inevitably leads us to consider
conifold configurations.

For most weighted hypersurfaces the 1--dimensional family just described
is only part of a higher--dimensional moduli space. The fractional
transformation mirror transform, of course,
 applies to mirror pairs involving larger moduli spaces as well.
An example that has received attention recently is given by the
family
\beq
\IP_{(1,1,2,2,2)}[8]\ni
\{p=z_1^8+z_2^8+z_3^4+z_4^4+z_5^4-8\psi \prod_{i=1}^5z_i -2\l z_1^4z_2^4 =0\}.
\eeq
This theory has the spectrum $(\ags,\gens)=(2,86)$ and the mirror can
be obtained by applying the fractional transformations
\beq
z_1=y_1^{7/8},~~z_2=y_1^{1/8}y_2^{3/4},~~z_3=y_2^{1/4}y_3^{3/4},~~
z_4=y_4^{1/4}y_5,~~z_5=y_5
\eeq
to the orbifold $\IP_{(1,1,2,2,2)}[8]/\ZZ_8\times \ZZ_3^2$ with respect to the
action
\beq
\ZZ_8 \times \ZZ_3^2:~~~
\left[\matrix{7&1&0&0&0\cr 0&3&1&0&0\cr 0&0&3&1&0\cr}\right].
\eeq
This leads to the polynomial
\beq
p= y_1^7+y_1y_2^6+y_2y_3^3+y_3y_4^4+y_5^4 - 8\psi \prod_{i=1}^5 y_i -2\l
y_1^4y_2^3
\eeq
which lives in the configuration $\IP_{(4,4,8,5,7)}[28]$ for which one
finds the mirror spectrum $(\ags,\gens)=(86,2)$, as expected.

Other examples which have been the focus of recent investigations of
the conjectured heterotic-type II duality
\cite{kv95,klm95,bcdffrsp95,klt95,vw95,agnt95,kklmv95,gc95}
can be analyzed in the same manner. The simpler of the two most
prominent members is the two--parameter family
\beq
\IP_{(1,1,2,2,6)}[12]^{(2,128)} \ni
\{p=z_1^{12}+z_2^{12}+z_3^6+z_4^6+z_5^2-12\psi \prod_{i=1}^5z_i
           -2\l z_1^6z_2^6
=0\}
\eeq
which is mapped via the $\ZZ_6^2\times \ZZ_2$ fractional transformations
$z_1=y_1$,$z_2=y_2^{5/6}$,$z_3=y_2^{1/6}y_3^{5/6}$,
$z_4=y_4^{1/6}y_5^{1/2}$ into the mirror family
\beq
\IP_{(25,30,54,82,109)}[300]\ni
\{p=y_1^{12}+y_2^{10}+y_2y_3^6+y_3y_4^3+y_4y_5^2-
12\psi \prod_{i=1}^5y_i -2\l y_1^6y_2^5 =0\},
\eeq
which describes the type IIB dual of a heterotic string vacuum with
129 hypermultiplets and 3 vector multiplets.
Finally, the three--parameter family
\beq
\IP_{(1,1,2,8,12)}[24]\ni
\{p=z_1^{24}+z_2^{24}+z_3^{12}+z_4^3+z_5^2-12\psi \prod_{i=1}^5z_i
              -2\l z_1^6z_2^6z_3^6 - \si z_1^{12}z_2^{12} =0\}
\eeq
and its mirror, obtained via the
$\ZZ_{24}\times \ZZ_3\times \ZZ_2$ fractional transformations
$z_1=y_1^{23/24}$,~$z_2=y_1^{1/24}y_2^{1/2}$,~$z_3=y_3^{2/3}$,
{}~$z_4=y_3^{1/3}y_4$ ,~ $z_5=y_2^{1/2}y_5$ into the mirror family
\beq
\IP_{(24,44,69,161,254)}[552] \ni
\{p=y_1^{23}+y_1y_2^{12}+y_2y_5^2+y_3^8+y_3y_4^3 -12\psi \prod_{i=1}^5 y_i
          -2\l y_1^6y_2^3y_3^4 - \si y_1^{12}y_2^6=0\}
\eeq
describe the type IIA and IIB duals respectively of a heterotic vacuum with
244 hypermultiplets and 4 vector multiplets.
Our construction clearly generalizes to ever more general moduli spaces.

\vskip .2truein
\noindent
{\bf 4. New Directions in the Global Calabi-Yau Moduli Space via General
        Conifold Transitions}

\noindent
The splitting construction in either the ordinary
projective class of Calabi-Yau manifolds \cite{cdls88}, or in the weighted
category, which we will discuss below, is particularly simple because
it provides simple representations of the different smooth phases that are
connected via a singular variety. In general such a simple description of
the manifold `on the other side' is not to be expected, even if one
starts out with a complete intersection. A class of manifolds which
illustrates the necessity of considering more general conifold
transitions is provided by the 1--parameter families of (\ref{1-dim-fam}).
A particularly simple subclass is obtained by considering
spaces of Brieskorn--Pham type
\beq
\IP_{(k_1,...,k_5)}[d] \ni
\left\{ \sum_{i=1}^5 k_i z_i^{d/k_i} - d\psi \prod_{i=1}^5 z_i =0\right\}.
\eeq
These varieties acquire singularities at $\psi^d=1$, the singular points are
nodes, and there are $d^3/\prod_i k_i$ of them.

The natural question arises what the manifolds are that are found after
traversal of the conifold. In the present case the splitting
construction does not provide insight and one has to take recourse to more
general considerations concerning the resolution of singularities in
Calabi-Yau manifolds. The general theory is rather more involved because it
is not automatically guaranteed that the resolved variety is
projective \cite{jw88}, in contrast to the splitting construction.
Once this question has been answered, however, it is not difficult
to compute the Hodge numbers. The nodes at the conifold point are resolved by
introducing a sphere $\IP_1 \sim S^2$, in contrast to blowing up. Thus
the surgery involves replacing a three-sphere $S^3$ by a projective curve,
thereby changing the Euler number by $+2$. Hence the Euler number
of the resolved manifold becomes\fnote{6}{For general weighted
    Calabi-Yau manifold this result is not
        correct as we will discuss in the last Section.}
\beq
\chi(\tM) = \chi(M) + 2N
\eeq
if $N$ is the number of nodes. In a Calabi-Yau manifold this can only be
achieved by increasing $h^{(1,1)}$ by unity or decreasing $h^{(2,1)}$
by unity. Thus
\beq
h^{(1,1)}(\tM) = h^{(1,1)}(M) + \delta,~~~~
h^{(2,1)}(\tM) = h^{(2,1)}(M) - (N-\delta)
\eeq
where $\delta$ is the number of linearly dependent vanishing cycles.

Consider e.g. the simplest space of Brieskorn--Pham type,
the family of quintics
(\ref{quint-def}) at $\psi^5 =1$.  The starting point here is a family
of smooth manifolds with $(h^{(1,1)}, h^{(2,1)}) = (1,101)$ which acquires
$d^3/\prod k_i =125$ nodes at the conifold. Thus the Euler number of the
resolved
manifold is $\chi(\tM) =-200+ 2\cdot 125 =50$ and the Hodge numbers are
$(h^{(1,1)}(\tM), h^{(2,1)}(\tM))=(1+\delta, 101-(125-\delta))$.
If we wish to fix the configuration
of the nodes then we expect the resolved manifold to have fewer complex
deformations since the resolution only introduces $\IP_1$s and we lose
the complex deformations which would kill the nodes. The quintic is rather
special since the number of  nodes it acquires at the conifold is larger
than the number of complex deformations one starts out with.
Therefore it leads to a resolved space which is rigid. Because there
are 24 more nodes than there are complex deformations one finds
$\delta =24$ and the  resolved space in fact has the Hodge numbers
$(h^{(1,1)}(\tM), h^{(2,1)}(\tM))=(25,0)$.
It has been checked in \cite{cs86} that the manifold is indeed Calabi-Yau.

A further example of a Brieskorn--Pham type variety whose conifold
transition leads to a rigid manifold as well is the one--parameter family
of hypersurfaces
\beq
\IP_{(1,1,1,1,2)}[6] \ni
\left\{\sum_{i=1}^4 z_i^6 + 2z_5^3 - 6\psi \prod_{i=1}^5 z_i =0\right\},
\eeq
which acquires a conifold configuration at $\psi=1$ with 108 nodes.
The resolution of these nodes leads to a smooth rigid
manifold with $\chi=12$.

The fact that the cohomology of this example is produced neither by the class
of all complete intersection Calabi-Yau manifolds \cite{cdls88} nor by the
class of all weighted hypersurfaces or, more generally, the complete class of
Landau--Ginzburg theories, shows that general resolutions allow us to explore
new, yet uncharted, territory of the global moduli space of all
Calabi-Yau manifolds.

\vskip .2truein
\noindent
{\bf 5. Calabi-Yau Isomorphisms: Connecting Collective Webs and new
        K3 fibrations.}

\noindent
In this Section we discuss two further applications of
fractional transformations
which will turn out to be useful in the following parts of the paper.
The first is that they lead to a particularly simple
class of intersection points between the moduli spaces of different types of
Calabi-Yau spaces, whereas the second shows how insight into the fiber
structure of certain Calabi-Yau manifolds can be gained from fractional
transformations.

It was shown in \cite{rs89} that the moduli space of Calabi-Yau
manifolds embedded in weighted projective spaces is connected to the moduli
space of manifolds embedded in products of ordinary projective space. This
arose simply because there exist
isomorphisms between weighted hypersurfaces and ordinary complete
intersections of higher
codimension, the simplest example being the relation
\beq
\IP_{(1,1,2,2,2)}[8]^{(2,86)} ~\sim~
\matrix{\IP_1\cr \IP_4\cr} \left[\matrix{2&0\cr 1&4\cr}\right].
\lleq{oct}
Fractional transformations in fact lead to an explanation simpler and more
general than the analysis of \cite{rs89}, providing a great many of such
identifications.
Consider the following class of manifolds of Brieskorn--Pham type
\beq
\IP_{(2k_1-1,2k_1-1,2k_2,2k_3,2k_4)}[2k]
\lleq{hyp-surfs}
with $k=(2k_1+k_2+k_3+k_4-1)$ and $2k/(2k_1-1) \in 2\IN$.
Viewing these string vacua as a Landau--Ginzburg theory we can add trivial
factors $y_i^2$ without changing the model. Adding two such factors
and applying the basic
isomorphism (\ref{basic-iso}) to the two parts $(x_i^{2k/(2k_i-1)}+y_i^2)$
in the resulting representation of the theory changes the configuration to
\beq
\IC_{(2(2k_1-1),2(2k_1-1),2k_2,2k_3,2k_4,(k_2+k_3+k_4),(k_2+k_3+k_4)}[2k].
\eeq
If the $\ZZ_2$'s happen to act trivially we can use
the construction of \cite{rs93} to derive the corresponding manifold
of codimension 2, arriving at the relations
\beq
\IP_{(2k_1-1,2k_1-1,2k_2,2k_3,2k_4)}[2k]
{}~\sim ~\matrix{\IP_{(1,1)}\hfill \cr
        \IP_{((2k_1-1),(2k_1-1),k_2,k_3,k_4)}\cr}
\left[\matrix{2&0\cr (2k_1-1) & k\cr}\right].
\lleq{equiv}
For $k_1=k_2=k_3=k_4=1$ we recover the example above,  discussed in
detail in \cite{rs89}. This class of spaces thus provides a great many
of identifications between hypersurfaces and Calabi-Yau manifolds of
higher codimension.

A second reason why this class of manifolds is of interest comes from the
fact that the relations (\ref{equiv}) are also useful for explorations
of spacetime mirror symmetry. It has been recognized early on \cite{klm95}
that important for the heterotic-type II duality \cite{kv95,fhsv95} is the
fact that the Calabi-Yau manifolds involved are K3--fibrations.
It is therefore of some importance to gain insight into the nature of
such manifolds in order to obtain further examples of dual pairs beyond
the few which have been the focus
of most of the discussions so far. Following the analysis of
\cite{cdfkm94} it can readily be seen that all manifolds
of the type (\ref{hyp-surfs}) are in fact K3 fibrations. Defining a divisor
$D_{\l} \in \IP_{(2k_1-1,2k_1-1,2k_2,2k_3,2k_4)}[(2k_1-1)~2k]$ via
$(z_1-\l z_2)=0$, and applying the (1--1) coordinate transformation
$y_1=z_1^2$, shows that the fibers are described by the K3 configurations
\beq
\IP_{(2k_1-1,k_2,k_3,k_4)}[k].
\lleq{k3-hyp}
This class thus provides a pool of K3 fibrations which considerably extends
the list of examples of K3 fibrations enumerated in \cite{klm95}.
For convenience
we provide the complete set of models in the Appendix.

An important aspect of the class of fibrations
(\ref{hyp-surfs}) is that the equivalences
(\ref{equiv}) trivially allow the identification of the (possible)
type II image of the heterotic dilaton.
This is because for N$=$2 heterotic vacua the dilaton couples to
the rest of the moduli in such a way
\cite{old-coup}\fnote{7}{See \cite{bcdffrsp95,dil-coup} for recent
reviews of this subject as well as a more complete list of
   the original references.}
that the intersection numbers of the corresponding modes on the
type II Calabi-Yau dual, denoted by $s$ and $m_i$, $i=1,...,n$,
take the form
\beq
\k_{sss}=0=\k_{ssi}, ~~~\k_{sij}={\rm diag}(1,n).
\eeq
This condition merely indicates that the corresponding Calabi-Yau dual
is a fibered manifold and leaves open a number of different ways to
fiber the manifold \cite{bh90}.
The condition derived from the heterotic theory
which identifies the fibers as K3 varieties is the fact that the dual
Calabi-Yau manifolds also have to satisfy $\int c_2(M) h_s=24$, where
$h_s$ is the element in H$^2(M)$ describing the dual image of the
dilaton\fnote{8}{We are grateful to B.Hunt for correspondence on
    this point. A more detailed recent discussion of these facts can be
   found in \cite{al95}.}.
For manifolds with large Picard number $b_2$ it is quite involved
to compute these couplings and identify the appropriate $h_s$.
For our class of fibrations (\ref{hyp-surfs}) however the equivalent
representation as a codimension--two space allows for an
immediate identification -- the image of the dilaton must be the
K\"ahler form which descends down to the Calabi-Yau space from the
ambient projective curve.

It should be noted that we have assumed the condition
$(2k/(2k_1-1))\in 2\IN$ for convenience of presentation only. It is
not necessary either for relations of the type we have discussed or for
the manifold to be a K3 fibration. An example which illustrates this
point is provided by the manifold
\beq
\IP_{(3,3,4,4,14)}[28]
{}~\sim ~\matrix{\IP_{(1,1)}\hfill \cr
        \IP_{(3,3,2,2,7)}\cr}
\left[\matrix{2&0\cr 3&14\cr}\right]
\eeq
with K3 fiber $\IP_{(3,2,2,7)}[14]$.
In this more general class some of the new heterotic spectra found in
\cite{afiq95} can be found and therefore it provides the
`missing' Calabi-Yau dual candidates of some known heterotic N=2 vacua.

A similar discussion applies to relations of the type
\beq
\IP_{(2k_1-1,2k_1-1,2k_2,2k_3,2k_4,2k_5)}[2a~~2b]
{}~\sim ~\matrix{\IP_{(1,1)}\hfill \cr
        \IP_{((2k_1-1),(2k_1-1),k_2,k_3,k_4,k_5)}\cr}
\left[\matrix{2&0&0\cr (2k_1-1) &a &b\cr}\right]
\lleq{2equiv}
where $(a+b)=2k_1-1+\sum_{i=2}^5k_i$. This class describes K3 fibrations
as well and generalizes the second type of Calabi-Yau spaces considered
in \cite{klm95}.

\vskip .2truein
\noindent
{\bf 6. Splitting and Contraction for Weighted CICYs and Spacetime Mirror
Symmetry}

\noindent
Singularities are ubiquitous in the moduli space of Calabi-Yau
spaces: no matter from which smooth point one starts, moving along a generic
complex deformation will eventually lead to a singular configuration.
What is not ubiquitous is knowledge about what happens `on the other side' of
the singularity, or whether it exists at all. The existence problem is
far from obvious since the projectivity of the small resolution of
singularities, obtained by deforming a family of smooth varieties $V_t$
into a singular configuration $V_0$, is not easy to check in general.

In ref. \cite{cdls88} a certain type of conifold transition between
Calabi-Yau spaces has been introduced which avoids this difficulty.
The constructions of \cite{cdls88}, called splitting and contraction, have
the virtue that they describe conifold transitions of the family
$V_t$ of (quasi--)smooth varieties depending on
some complex variable $t$
\beq
V_t \lra V_0 \lra \tV
\lleq{gen-spl}
which automatically provide relations between smooth {\it Calabi-Yau}
manifolds (here $\tV$ denotes a small resolution Calabi-Yau manifold). We
will show in this Section that the construction of \cite{cdls88}
generalize to the weighted framework even though the story acquires some
new twists.

There first new constraint that is specific to the class of weighted
manifolds and has no counterpart in the ordinary projective class originates
from the fact that for a given choice of weights
there may not exist a quasismooth set of polynomials
\fnote{9}{We will not discuss possible generalizations, such as the one
      discussed in \cite{cdk95}, in the present paper.}.
The problem is even more pronounced in the case of complete intersections
with higher codimension than it is for hypersurfaces in weighted projective
four--space, as discussed in the first reference of \cite{ms},
and has in fact been one of the major stumbling blocks for the
construction of the class of all Calabi-Yau manifolds embedded
in products of weighted projective
spaces. To illustrate the problem consider the following split
\beq
\IP_{(k_1,k_1,k_2,k_3,k_4)}[d] \lra
\matrix{\IP_{(1,1)} \hfill \cr \IP_{(k_1,k_1,k_2,k_3,k_4)}\cr }
\left[ \matrix{1&1\cr ak_1&(d-ak_1)\cr}\right] \ni
\left\{\begin{tabular}{l}
         $p_1= x_1Q(y_i) + x_2R(y_i)$ \tabroom \\
         $p_2 = x_1S(y_i) + x_2T(y_i)$ \tabroom \\
       \end{tabular}
\right\},
\eeq
where $d=2k_1+k_2+k_3+k_4$ and $a$ is some postive integer. For
$k_1=k_2=k_3=k_4=1$ this reduces to the simplest type of split
considered in \cite{cdls88}, the rhs describing a $\IP_1$--split $\tV$ of the
determinantal variety
\beq
\IP_{(k_1,k_1,k_2,k_3,k_4)}[d] \ni V_0 =\{p=QT-RS=0\},
\eeq
which can be deformed into a smooth variety $V_t$ (for favourable choices of
weights).

Now if, for instance, the weights are such that the
first polynomial involves only the first two coordinates of the weighted
4--space, then it is never possible to find transverse choices of polynomials.
The equations that follow from the transversality condition, according to which
$dp_1\wedge dp_2=0$ may not have any solution on the manifold, lead to two
branches. It suffices to discuss one of these. Assuming that indeed
$Q=Q(y_1,y_2)$ and $R=R(y_1,y_2)$ leads to
$0\equiv Q{\Big |}_{(0,0,y_3,y_4,y_5)}$ and
$0\equiv R{\Big |}_{(0,0,y_3,y_4,y_5)}$, and therefore the
equations restricted to the subvariety parametrized by
$(0,0,y_3,y_4,y_5)$ reduce to $0=T$ and
\beq
\left(S\frac{\del R}{\del y_i}\right){\Big |}_{(0,0,y_3,y_4,y_5)}~= 0 =
\left(\frac{\del R}{\del y_i} \frac{\del T}{\del y_j}
       - \frac{\del R}{\del y_j} \frac{\del T}{\del y_i}\right)
        {\Big |}_{(0,0,y_3,y_4,y_5)}
\eeq
for all $i$ and all $i<j$ respectively.
If  $a>1$ then $\del R/\del y_i{\Big |}_{(0,0,y_3,y_4,y_5)} \equiv 0$
and the configuration is singular for all points on the
curve $\IP_{(k_2,k_3,k_4)}[d-ak_1]$. A codimension two
Calabi-Yau configurations for which it is not possible to find quasismooth
choice of polynomials is given by $(k_1,k_2,k_3,k_4)=(1,3,3,3)$ with
$a=2$, for instance. Assuming, then, that $a=1$ the analysis of the
transversality equations reveals that quasismoothness can be obtained
by requiring that both  $(\del S/\del y_1)=0$ and $(\del T/\del y_2)=0$
and that both, $S$ and $T$, depend on all but at most one variables, and
that they are of standard type in these variables. Furthermore, if $S$
is independent of some variable then the polynomial $T$ must be of Fermat
type in this variable, and vice versa.

A simple example of a splittable configuration is provided
by  the quasismooth octic hypersurface
\beq
M=\IP_{(1,1,2,2,2)}[8]^{(2,86)} \ni
      \left\{\sum_i z_i^8 + \sum_j z_j^4 =0\right\}.
\lleq{octic}
This manifold can be split into the codimension-two variety
\beq
M_{\rm split} =\matrix{\IP_{(1,1)} \hfill \cr \IP_{(2,2,1,1,2)}\cr }
\left[ \matrix{1&1\cr 2&6\cr}\right] \in
\left\{\begin{tabular}{l}
        $p_1= x_1y_1 + x_2y_2$ \tabroom \\
        $p_2=x_1(y_2^3+y_4^6 -y_5^3) +  x_2(y_1^3+y_3^6+y_5^3)$ \tabroom \\
       \end{tabular}
\right\},
\lleq{split-oct}
which can be checked to be transverse. The determinantal variety leads to
the singular octic
\beq
p_s = Q(y_i)T(y_i) - R(y_i)S(y_i)
 = y_1 (y_1^3+y_3^6+y_5^3) - y_2 (y_2^3+y_4^6 -y_5^3)
\eeq
fails to be transverse at $\IP_{(2,2,1,1,2)}[2~2~6~6] = 18$ nodes.
Hence the Euler number of the codimension two complete intersection is
$\chi(M_{\rm split}) = \chi(M) +2\cdot 18 = -168 + 36 = -132$.
Since $h^{(1,1)}=3$, because of the additional $\IP_1$,  the complete
massless spectrum that results is $(h^{(1,1)}, h^{(2,1)}) = (3,69)$.

The conifold transition
\beq
\IP_{(1,1,2,2,2)}[8]^{(2,86)} ~\lolra ~
\matrix{\IP_{(1,1)} \hfill \cr \IP_{(2,2,1,1,2)}\cr }
\left[ \matrix{1&1\cr 2&6\cr}\right]^{(3,69)}
\lleq{oct-split}
is of some interest because it provides a possible ingredient of a
sequence of string spectra discovered by Kachru and Vafa in their
discussion of dual pairs of type II Calabi-Yau compactifications
and heterotic K3$\times $T$^2$ vacua. Starting with the
 E$_8\times $E$_8$ heterotic string they considered a series of
embeddings of
SU(N) factors into one of the E$_8$s, thereby breaking this
group down to E$_7$ (for N$=$2), E$_6$ (N$=$3), SO(10) (for N$=$4), or
SU(5) (for N$=$5), respectively. The spectra obtained in this way are
\cite{kv95}
\beq
{\rm N}=2:~(65,19),~~~{\rm N}=3:~(84,18),~~~{\rm N}=4:~(101,17),~~~
{\rm N}=5:~(116,16).
\lleq{sh-cu}
This sequence is intriguing: it has precisely the structure
we expect from splitting transitions of the type discussed above
\fnote{10}{Not of a general conifold transition however, as follows from
          our discussion in Section 4.}:
because of the vanishing cycles the number of complex deformations is
reduced in the transition $V_t \lra V_0$ and because
of the properties of small resolutions new K\"ahler deformations
are introduced in the smoothing process $V_0 \lra \tV$. However, because
the purported Hodge numbers $(h^{(1,1)},h^{(2,1)})$ do not appear for
any of the manifolds in the list of all CICYs \cite{cdls88}
nor for any of the models in the list of all Landau--Ginzburg
theories \cite{all-lgs}, it seems difficult at present to check for the
possibility of a direct, simple analog of the splitting transition.
Kachru and Vafa, however, made the intriguing observation that the
transition from N$=$5 to N$=$4 is reminiscent of the splitting process
of \cite{cdls88} applied to the codimension two Calabi-Yau manifold
\beq
\IP_4[5]^{(1,101)} ~\lolra ~
 \matrix{\IP_1 \hfill \cr \IP_4\cr }
\left[ \matrix{1&1\cr 1&4\cr}\right]^{(2,86)},
\lleq{quint-split}
provided an overall shift of 14 in the Hodge numbers is
taken into account\fnote{11}{The origin of this shift remains
         obscure at present.}.
This shift of 14 results in the sequence of Euler numbers
\beq
{\rm N}=2:~-92,~~~{\rm N}=3:~-132,~~~{\rm N}=4:~-168,~~~ {\rm N}=5:~-200.
\lleq{euler-seq}
The idea that there might indeed exist an analog of Calabi-Yau
splitting in the context
of N$=$2 heterotic string theory clearly would gain support if direct splits
could be found for the remaining two embeddings of SU(N).
We see that a candidate for the second element in the chain
(\ref{euler-seq}) is provided by the pair of spaces connected through the
weighted split (\ref{oct-split}). To find the remaining
elements of the sequence (\ref{euler-seq}), recall from Section 5 that
the octic $\IP_{(1,1,2,2,2)}[8]$ has another representation as a
codimension-two ordinary complete intersection Calabi-Yau manifold.
Since we have just found the split of the quasismooth octic to a
(3,69) manifold we might expect that an appropriate direct split of
the second representation might exist as well. Indeed, using the
ordinary splitting of \cite{cdls88} we find the sequence of splits
\beq
\matrix{\IP_1\cr \IP_4\cr} \left[\matrix{2&0\cr 1&4\cr}\right]_{-168}
{}~\lolra ~
\matrix{\IP_1\cr \IP_1\cr \IP_4\cr}
           \left[\matrix{0&1&1\cr 2&0&0\cr 1&1&3\cr}\right]_{-132}
{}~\lolra ~
\matrix{\IP_1\cr \IP_1\cr \IP_2\cr \IP_4\cr}
   \left[\matrix{1&1&0&0&0\cr 0&0&2&0&0\cr 1&0&0&1&1\cr
0&1&1&1&2\cr}\right]_{-92}.
\lleq{split-seq}
Jumping ahead we emphasize that all manifolds of
(\ref{split-seq}),(\ref{split-oct}) are K3 fibrations.
 Thus we have established a direct split within the subclass of
K3 fibered Calabi-Yau manifolds for each element in the sequence of
SU(N)--embeddings discussed by Kachru and Vafa. The ability to find
direct splits according to growing N may be interpreted as
evidence that there might exist an alternative construction of Calabi-Yau
splitting in the context of  N$=$2 theories.

At present no conifold transition between Calabi-Yau manifolds
with heterotic duals is known. As an initial step in this
direction we present the first conifold transition of a
Calabi-Yau manifold the dual of which has been identified by
Kachru and Vafa\fnote{12}{This problem is also under consideration in the
   work of \cite{bkk95}}.
The heterotic vacuum in question is constructed by
starting with an eight--dimensional compactification with an enhanced
gauge group E$_8 \times $E$_8\times$SU(3)$\times $U(1)$^2$, obtained
by going to a special point in the moduli space of the torus T$^2$.
Choosing particular embeddings of the various relevant bundles into
the gauge group factors one ends up with a theory with 102 hypermultiplets
and 6 vector multiplets. Since there is only one space in the class
(\ref{equiv}) with the appropriate spectrum, given by
$\IP_{(1,1,2,4,4)}[12]^{(5,101)}$, it is very likely that this indeed
describes the dual of the heterotic dual just described.
This configuration can be split as follows
\beq
\IP_{(1,1,2,4,4)}[12] ~\lolra ~
\matrix{\IP_{(1,1)}\hfill \cr \IP_{(4,4,1,1,2)}\cr}
\left[\matrix{1&1\cr 4&8\cr}\right],
\lleq{hetsplit}
involving a conifold configuration with 32 nodes. The resulting
cohomology for the split manifold is
$(h^{(1,1)}, h^{(2,1)})=(6,70)$.
We will show in the next Section that this conifold transition belongs
to a whole class of weighted splits which connect Calabi-Yau manifolds
that are all K3 fibrations. Thus our split (\ref{hetsplit}) remains
in the class of spaces relevant for heterotic-type II duality.
Orbifolding this manifold by $\ZZ_6\times \ZZ_3^2$
and iterating the basic isomorphism we find the mirror configuration
to be $\IP_{(20,24,49,54,93)}[240]^{(101,5)}$.

At this time only
very few heterotic vacua along the lines of \cite{kv95} have been
constructed, with results \cite{afiq95} that are not
too different from the corresponding heterotic spectrum
$(n_H,n_V)=(71,7)$. It should be expected that our splitting result
will turn up as the number of heterotic vacua grows.

\vskip .2truein
\noindent
{\bf 7. Conifold Transitions between K3 fibered Calabi-Yau manifolds.}

\noindent
As a second application we show how weighted splitting indicates that the
deeper understanding recently obtained \cite{klm95} of the appearance of the
$j$--function in the context of spacetime mirror symmetry \cite{kv95,fhsv95}
is far more general than initially thought. In order to do so, we
recall that the underlying reason for the appearance of the $j$--function
is to be found in the K3 fibration of the
Calabi-Yau threefold. Because the web of Calabi-Yau manifolds can be
traversed via conifold transitions, it is natural to ask what the behavior of
K3 fibrations is under such transitions. Our discussion in the following
will focus on the splitting construction, and for simplicity we discuss in
detail one example, the quasismooth octic.

It is well--known that the octic (\ref{oct}) is a  K3 fibration
 \cite{cdfkm94}, i.e. the linear system L defined by the linear sections
defines a family of K3--surfaces in the representation $\IP_3[4]$.
In more detail consider the divisor defined by the linear relation
$z_2 = \th z_1$, which leads to the family of hypersurfaces
\beq
\IP_{(1,2,2,2)}[8] \ni
\{p=(1+\th^8-2\l \th^4)z_1^8+z_3^4+z_4^4+z_5^4-8\psi \th
z_1^2\prod_{i=3}^5z_i=0\}.
\eeq
With  $y_1=z_1^2$ and $y_i=z_{i+1}, i=2,3,4$, one arrives at the family
of quartic K3--hypersurfaces of the form
\beq
\IP_3[4]\ni \{p=(1+\th^8-2\l \th^4)y_1^4+y_2^4+y_3^4+y_4^4-8\psi \th
\prod_{i=1}^4y_i =0\}.
\eeq
It is this structure of the K3 fibration of the Calabi-Yau threefold which
explains \cite{klm95} the appearance of the $j$--function \cite{kv95}.

Now, starting from the codimension two split of the octic as defined in
(\ref{oct-split}), contraction leads to the determinantal variety
\beq
\IP_{(1,1,2,2,2)}[8] \ni
\{p_0 \equiv y_1(y_1^3 + y_3^6-y_5^3)-y_2(y_2^3+y_4^6+y_5^3)=0\}
\eeq
which, via $y_4=\th y_3$, and the definitions $z_i=y_i, i=1,2$, $z_3=y_3^2$,
$z_4=y_5$, leads to the family of singular K3 surfaces
\beq
\IP_3[4] \ni \{z_1^4-z_2^4+(z_1-\th z_2)z_3^3 +(z_1-z_2)z_4^3=0\}.
\lleq{det-k3}
Furthermore the codimension two variety describing the split contains
the family of K3 surfaces
\beq
\matrix{\IP_1 \cr \IP_3\cr} \left[\matrix{1&1\cr 1&3\cr}\right] \ni
\left\{ \begin{tabular}{l}
          $p_1=x_1y_1 + x_2y_2$ \tabroom \\
 $p_2= x_1(y_2^3+\th y_3^3+y_4^3) + x_2(y_1^3+y_3^3-y_4^3)$  \tabroom \\
         \end{tabular}
\right\}
\eeq
which can be seen to lead precisely to the determinantal K3 of (\ref{det-k3}).
Thus we see that the splitting and contraction process not only relates
K3--fibration but essentially takes place in the fiber, passing through
singular K3 surfaces. The splitting conifold transitions therefore carry
over  the K3 fiber structure of the hypersurfaces to more
complicated Calabi-Yau manifolds of higher codimension.

The above analysis clearly allows for generalizations. A simple class
of splits is defined as follows
\beq
\IP_{(2k-1,2k-1,2l,2l,2m)}[2(2k-1+2l+m)]~ \lolra ~
\matrix{\IP_{(1,1)}\hfill \cr \IP_{(2l,2l,2k-1,2k-1,2m)}\cr}
\left[\matrix{1&1\cr 2l&2(2k-1+l+m)\cr}\right],
\eeq
where the codimension one hypersurfaces, containing the K3 surfaces
$\IP_{(2k-1,l,l,m)}[2k-1+2l+m]$, split into codimension two manifolds
containing codimension-two K3 manifolds
\beq
\IP_{(2k-1,l,l,m)}[2k-1+2l+m] ~\lolra ~
\matrix{\IP_{(1,1)}\hfill \cr \IP_{(l,l,m,2k-1)}\cr}
\left[\matrix{1&1\cr l&(l+m+2k-1)\cr}\right].
\eeq
As in the example above the quasismooth K3 hypersurfaces on the lhs
are deformations of the determinantal K3s obtained by contracting the
codimension-two K3 complete intersections of the split manifold.
This construction is not restricted to this simple manifolds but
can also be applied also to a variety of other classes, starting from
more complicated spaces of higher codimension such as
\beq
\IP_{(2k-1,2k-1,2l,2l,2m,2n)}[2a~~2b]~ \lolra ~
\matrix{\IP_{(1,1)}\hfill \cr \IP_{(2l,2l,2k-1,2k-1,2m,2n)}\cr}
\left[\matrix{1&1&0\cr 2l&2(a-l)&2b\cr}\right],
\eeq
with $(a+b)=2k-1+2l+m+n$, or, more concretely,
\beq
\matrix{\IP_{(1,1)}\hfill \cr \IP_{(1,1,1,1,2,2)}\cr}
\left[\matrix{1&1&0\cr 2&2&4\cr}\right]~\lolra ~
\matrix{\IP_{(1,1)}\hfill \cr \IP_{(1,1)}\hfill \cr
        \IP_{(1,1,1,1,2,2)}\cr}
\left[\matrix{0&0&1&1\cr 1&1&0&0\cr 2&2&2&2\cr}\right],
\eeq
all of which are K3 fibrations.

The above splitting analysis provides strong evidence that the appearance
of the $j$--function in the analysis of the heterotic-type II duality
is not restricted to the simple Calabi-Yau spaces that have been
considered in the literature but instead
extends to the general class of weighted complete
intersection Calabi-Yau manifolds of arbitrary codimension.

\vskip .2truein
\noindent
{\bf 8. Colliding Singularities}

\noindent
Finally, we wish to point out a novel phenomenon that arises in
conifold transitions between  weighted Calabi-Yau manifolds. Namely,
it can happen that a number $N_i$ of the $N$ hypersurface singularities
sit on top of $\ZZ_{p_i}$ orbifold singularities of the weighted space.
If such a situation occurs the results obtained for conifold
transitions between manifolds embedded
in products of ordinary projective spaces \cite{cdls88} are no longer
correct.

Consider the manifold
\beq
M=\IP_{(2,2,3,3,5)}[15]^{(7,43)} \ni
\left\{z_1^6z_3 + z_2^6z_4 + z_3^5 + z_4^5 +z_5^3=0\right\}
\eeq
with the split configuration
\beq
M_{\rm split} = \matrix{\IP_{(1,1)} \hfill \cr \IP_{(3,3,2,2,5)}\cr }
\left[ \matrix{1&1\cr 3&12\cr}\right],
\eeq
a quasismooth manifold of which is defined by the polynomials
\bea
0&=&p_1 = x_1y_1 + x_2y_2 \nn \\
0&=&p_2 = x_1(y_2^4+y_4^6 +y_4y_5^2)+x_2(y_1^4+y_3^6+y_3y_5^2).
\eea
Because this configuration does not allow a Fermat type
choice for the polynomials $S$ and $T$ enumerating the
singularities involves the detailed structure of the defining polynomials.
The number of nodes in this case is given by
$N=\IP_{(2,2,3,3,5)}[3~3~12~12]=8$, and therefore we might have expected
that the split manifold has Euler number $-56$. Computing the Euler
number of the split manifold with the standard methods however leads
to $\chi(M_{\rm split}) = -48$. The resolution of this discrepancy is
found by noting that one of the nodes sits on top of a $\ZZ_5$
orbifold singularity. Hence the resolution
\beq
\chi(M_{\rm split}) = \chi(M) + 2N + \sum_i (p_i-1)N_i
\eeq
leads to an additional contribution of $+8$ in the naive result,
leading to agreement with the standard computation.

\vskip .2truein
\noindent
{\bf Acknowledgement}

\noindent
We are grateful to P. Berglund, S. Chaudhuri, S. Lang, J. Louis and
F. Quevedo for discussions, and in particular B.Hunt for correspondence.
This work was supported in part  by NSF grant PHY--94--07194.
M.L. thanks the Theory Group at the University of Bonn for
hospitality during the course of part of this work.

\vfill \eject

\noindent
{\bf Appendix:}~{\it The class of K3 fibrations of the type
     $\IP_{(2k_1-1,2k_1-1,2k_2,2k_3,2k_4)}[2k]$, with
     $k=2k_1-1+k_2+k_3+k_4$ and $2k/(2k_1-1) \in 2\IN$.}

\baselineskip=16pt
\begin{center}
\begin{scriptsize}
\begin{tabular}{| r r r r |}
\hline
$\chi$ &$h^{(1,1)}$  &$h^{(2,1)}$  &Weights \tabroom \\
\hline
\hline
 108&  60&   6  &(14,16,20,25,25) \\
  96&  59&  11  &(10,12,33,33,44) \\
  96&  55&   7  &(10,12,16,19,19) \\
  84&  54&  12  &(8,10,27,27,36) \\
  84&  54&  12  &(6,10,19,19,22) \\
  84&  50&   8  &(8,12,14,17,17) \\
  72&  49&  13  &(6,10,11,11,28) \\
  72&  49&  13  &(6,8,17,17,20) \\
  72&  44&   8  &(9,9,14,16,24) \\
  64&  43&  11  &(7,7,8,12,22) \\
  60&  44&  14  &(5,5,12,16,22) \\
  48&  35&  11  &(6,7,7,10,12) \\
  48&  43&  19  &(4,10,21,21,28) \\
  48&  41&  17  &(4,9,9,10,22) \\
  48&  39&  15  &(5,5,8,14,18) \\
  48&  39&  15  &(4,12,14,15,15) \\
  48&  39&  15  &(4,10,12,13,13) \\
  36&  38&  20  &(6,8,21,21,28) \\
  36&  38&  20  &(4,6,13,13,16) \\
  36&  34&  16  &(4,8,10,11,11) \\
  32&  33&  17  &(5,5,6,8,16) \\
  24&  32&  20  &(4,5,5,12,14) \\
  24&  29&  17  &(8,10,12,15,15) \\
  24&  27&  15  &(5,5,6,6,8) \\
  24&  33&  21  &(4,6,7,7,18) \\
  12&  36&  30  &(2,12,15,15,16) \\
  12&  36&  30  &(2,10,13,13,14) \\
  16&  31&  23  &(4,5,5,8,18) \\
   0&  35&  35  &(2,12,21,21,28) \\
   0&  27&  27  &(4,6,15,15,20) \\
   0&  34&  34  &(3,3,8,14,20) \\
   0&  23&  23  &(4,6,8,9,9) \\
   0&  23&  23  &(4,4,6,7,7) \\
   0&  31&  31  &(2,7,7,10,16) \\
   0&  31&  31  &(2,8,11,11,12) \\
 -12&  38&  44  &(3,3,8,20,26) \\
 -12&  30&  36  &(2,8,15,15,20) \\
 -12&  30&  36  &(2,6,11,11,14) \\
 -12&  25&  31  &(3,3,4,10,10) \\
 -24&  20&  32  &(3,3,4,4,10) \\
 -24&  29&  41  &(2,6,7,7,20) \\
 -24&  27&  39  &(3,3,4,10,16) \\
 -24&  23&  35  &(2,5,5,6,12) \\
 -36&  20&  38  &(2,4,7,7,8) \\
 -36&  20&  38  &(2,6,9,9,10) \\
 -48&  31&  55  &(3,3,4,16,22) \\
 -48&  15&  39  &(2,4,4,5,5) \\
 -48&  21&  45  &(2,4,5,5,14) \\
 -48&  19&  43  &(2,3,3,8,8) \\
 -48&  19&  43  &(2,4,9,9,12) \\
\hline
\end{tabular}
{}~~~~~~~~~~~~~~~~~~~~~
\begin{tabular}{| r r r r |}
\hline
$\chi$ &$h^{(1,1)}$  &$h^{(2,1)}$  &Weights \tabroom \\
\hline
\hline
\hline
 -64&  11&  43  &(6,7,7,8,28) \\
 -72&  21&  57  &(2,3,3,8,14) \\
 -72&  13&  49  &(5,5,8,12,30) \\
 -72&  10&  46  &(4,5,5,6,20) \\
 -72&   7&  43  &(4,5,5,6,10) \\
 -84&  12&  54  &(2,2,5,5,6) \\
 -96&  14&  62  &(3,3,8,10,24) \\
 -96&  11&  59  &(3,3,4,8,18) \\
 -96&  11&  59  &(2,2,3,3,8) \\
 -96&   7&  55  &(3,3,4,6,8) \\
 -96&   5&  53  &(2,3,3,4,6) \\
-108&   6&  60  &(2,2,2,3,3) \\
-112&   7&  63  &(2,5,5,8,20) \\
-120&  25&  85  &(2,3,3,14,20) \\
-120&   6&  66  &(2,3,3,4,12) \\
-132&   7&  73  &(3,3,6,8,10) \\
-168&   2&  86  &(1,1,2,2,2) \\
-192&  11& 107  &(3,3,4,20,30) \\
-192&   8& 104  &(1,1,4,4,6) \\
-192&   5& 101  &(1,1,2,4,4) \\
-192&   3&  99  &(1,1,2,2,4) \\
-204&  14& 116  &(3,3,8,28,42) \\
-204&   9& 111  &(1,1,4,6,6) \\
-232&   9& 125  &(1,1,4,6,8) \\
-232&   5& 121  &(1,1,2,4,6) \\
-240&  11& 131  &(1,1,6,8,8) \\
-240&   7& 127  &(2,3,3,16,24) \\
-252&   2& 128  &(1,1,2,2,6) \\
-264&  11& 143  &(1,1,6,8,10) \\
-272&   7& 143  &(1,1,4,4,10) \\
-288&   9& 153  &(1,1,4,8,10) \\
-288&   4& 148  &(1,1,2,4,8) \\
-304&  12& 164  &(1,1,8,10,12) \\
-312&  11& 167  &(1,1,6,10,12) \\
-312&   8& 164  &(1,1,4,6,12) \\
-312&   5& 161  &(1,1,2,6,8) \\
-348&  12& 186  &(1,1,8,12,14) \\
-368&  10& 194  &(1,1,6,8,16) \\
-372&   8& 194  &(1,1,4,8,14) \\
-372&   4& 190  &(1,1,2,6,10) \\
-420&  10& 220  &(1,1,6,10,18) \\
-432&  13& 229  &(1,1,12,16,18) \\
-432&  11& 227  &(1,1,8,10,20) \\
-480&  11& 251  &(1,1,8,12,22) \\
-480&   3& 243  &(1,1,2,8,12) \\
-528&   7& 271  &(1,1,4,12,18) \\
-612&  12& 318  &(1,1,12,16,30) \\
-624&   9& 321  &(1,1,6,16,24) \\
-732&  10& 376  &(1,1,8,20,30) \\
-960&  11& 491  &(1,1,12,28,42) \\
\hline
\end{tabular}
\end{scriptsize}
\end{center}

\baselineskip=17.5pt
\vfill \eject


\begin{thebibliography}{9}
\bibitem{cdls88} P.Candelas, A.Dale, C.A.L\"utken and R.Schimmrigk,
                   Nucl.Phys. {\bf B298}(1988)493
\bibitem{as95} A.Strominger, Nucl.Phys. {\bf B451}(1995)96, hep-th/9504090
\bibitem{ms} P.Candelas, M.Lynker and R.Schimmrigk,
           Nucl.Phys. {\bf B341}(1990)383;
          B.R.Greene and R.Plesser, Nucl.Phys. {\bf B338}(1990)15
\bibitem{kv95} S.Kachru and C.Vafa, Nucl.Phys. {\bf B450}(1995)69,
        hep-th/9505105;
\bibitem{fhsv95} S.Ferrara, J.A.Harvey, A.Strominger and C.Vafa,
       hep-th/9505162
\bibitem{gms95}  B.R.Greene, D.Morrison and A.Strominger,
        Nucl.Phys. {\bf B451}(1995)109, hep-th/9504145
\bibitem{gh88} P.Green and T.H\"ubsch, Commun.Math.Phys. {\bf 119}(1988)431;
                Phys.Rev.Lett. {\bf 61}(1988)1163
\bibitem{cgh90} P.Candelas, P.Green and T.H\"ubsch,
          Nucl.Phys. {\bf B330}(1990)49
\bibitem{ls90} M.Lynker and R.Schimmrigk, Phys.Lett. {\bf B249}(1990)237
\bibitem{pa95} P.Aspinwall, hep-th/9510142
\bibitem{klm95} A.Klemm, W.Lerche and P.Mayr, hep-th/9506112
\bibitem{al95} P.Aspinwall and J.Louis, hep-th/9510234
\bibitem{all-lgs} A.Klemm and R.Schimmrigk, Nucl.Phys. {\bf B411}(1994)559,
                hep-th/9204060;
                  M.Kreuzer and H.Skarke, Nucl.Phys. {\bf B388}(1992)113,
                hep-th/9205004
\bibitem{bcdffrsp95} M.Bill\'{o}, A.Ceresole, R.D'Auria, S.Ferrara,
    P.Fr\'{e}, T.Regge, P.Soriani and A.van Proeyen, hep-th/9506075
\bibitem{klt95} V.Kaplunovsky, J.Louis and S.Theisen, hep-th/9506110
\bibitem{vw95} C.Vafa and E.Witten, hep-th/9507050
\bibitem{agnt95} I.Antoniadis, E.Gava, K.S.Narain and T.R.Taylor,
    hep-th/9507115
\bibitem{kklmv95} S.Kachru, A.Klemm, W.Lerche, P.Mayr and C.Vafa,
            hep-th/9508155
\bibitem{gc95} G.Curio, hep-th/9509042; hep-th/9509142
\bibitem{jw88} F.Hirzebruch, Max Planck Institut--Bonn preprint SFB/MPI 85--58;
            J.Werner, Bonner Math. Schriften, Vol. 186, 1987
\bibitem{cs86} C.Schoen, J.Reine und Angew.Math. {\bf 364}(1986)85
\bibitem{rs89} R.Schimmrigk, Phys.Lett. {\bf B229}(1989)227
\bibitem{rs93} R.Schimmrigk, Phys.Rev.Lett. {\bf 70}(1993)3688,
       hep-th/9210062; hep-th/9405086
\bibitem{cdfkm94} P.Candelas, X.de la Ossa, A.Font, S.Katz and D.R.Morrison,
            Nucl.Phys. {\bf B416}(1994)481, hep-th/9308083
\bibitem{old-coup} E.Cremmer, C.Kounnas, A.Van Proeyen, J.P.Derendinger,
  S.Ferrara, B.de Wit and L.Girardello, Nucl.Phys. {\bf B250}(1985)385;
   B.de Wit, P.G.Lauwers and A.Van Proeyen, Nucl.Phys. {\bf B255}(1985)569
\bibitem{dil-coup} A.Ceresole, R.D'Auria, S.Ferrara and A.Van Proeyen,
     hep-th/9502072; B.de Wit, V.Kaplunovsky, J.Louis and D.L\"ust,
     hep-th/9504034
\bibitem{bh90} B.Hunt, J.f.Reine und Angew. Math. {\bf 411}(1990)137
\bibitem{cdk95} P.Candelas, X.de la Ossa and S.Katz,
         Nucl.Phys. {\bf B450}(1995)267, hep-th/9412117
\bibitem{afiq95} G.Aldazabal, A.Font, L.E.Ib\'{a}nez and F.Quevedo,
 hep-th/9510093
\bibitem{bkk95} P.Berglund, S.Katz and A.Klemm, in preparation
\end{thebibliography}
\end{document}